\documentclass[12pt]{article}           
\long\def\new#1\endnew{{\bf #1}}    %\def\draftmode{class ~ \rm~by~ \it m.k.}
\def\preprint{UTTG-07-98\\TUW--98-13}       
\def\finished{May 1998}
\def\archive {hep-th/9805190}           

\def\title{  Classification of Reflexive Polyhedra \\[5mm] in Three Dimensions}

\long\def\abstract{ 
We present the last missing details of our algorithm for the classification
of reflexive polyhedra in arbitrary dimensions.
We also present the results of an application of this algorithm to the
case of three dimensional reflexive polyhedra.
We get 4319 such polyhedra that give rise to K3 surfaces embedded in toric
varieties. 16 of these contain all others as subpolyhedra.
The 4319 polyhedra form a single connected web if we define two polyhedra
to be connected if one of them contains the other.
}
\def\cont#1{\mathop{\vtop{\ialign{##\crcr $\hfil\displaystyle
                     {#1}\hfil$\crcr\noalign{\kern3 pt \nointerlineskip}
                     $\bracelu\leaders\vrule\hfill\leaders\vrule\hfill
                     \braceru$\crcr\noalign{\kern3 pt }}}}\limits}

\def\CY{Calabi--Yau}
\def\Dm{\Delta_{\rm max}}
\def\vpm{vertex pairing matrix}
\def\ip{\hbox{\bf 1}}\def\ipo{\hbox{\bf 0}}

\input epsf

\usepackage{amssymb,latexsym}			\catcode`\"=\active \let"=\"

\def\ifundefined#1{\expandafter\ifx\csname#1\endcsname\relax}
\def\bye{\end{document}}   

\def\HS#1 {\hspace*{#1pt}} \def\VS#1 {\vspace*{#1pt}} \long\def\del#1\enddel{} 
\def\BC{\begin{center}}    
\def\EC{\end{center}}

\let\0=\over      \let\1=\vec      \def\2{{1\over2}}    \let\3=\ss
\let\4=\underline \let\5=\overline \let\6=\partial      \def\7#1{{#1}\llap{/}}
\def\8#1{{\textstyle{#1}}}         \def\9#1{{\ifmmode{\pmb{#1}}\else\bf#1\fi}}

          \def\({\left(}       \def\){\right)}

\def\eeql#1 {\label{#1}\eeq}        
\def\beq{\begin{equation}}      \def\eeq{\end{equation}}        
\def\bea{\begin{eqnarray}}      \def\eea{\end{eqnarray}} 

\def\mao#1{\mathop{\rm #1}\nolimits}  
              
\def\mod{\mao{mod}}

\let\and=\wedge

\let\bra=\langle        \let\ket=\rangle        \def\<#1\>{\bra #1 \ket}

\def\rel#1 #2{\buildrel #1 \over {#2}}  
\def\fnote#1#2{\begingroup\def\thefootnote{#1}\footnote{#2}
                \addtocounter{footnote}{-1}\endgroup}   

\def\subdef#1{\gdef\globalColor##1{##1}}      
      \subdef{Black}

         \let\th=\theta

             \let\D=\Delta

\def\ce{{\cal E}}

\def\IR{{\mathbb R}}  \def\IP{{\mathbb P}} 
   
\def\IZ{{\mathbb Z}} \def\IQ{{\mathbb Q}}

\def\plb#1 #2 {Phys. Lett. {\bf B#1} #2 }
\def\phr#1 #2 {Phys. Rep. {\bf  #1} #2 }        
\def\npb#1 #2 {Nucl. Phys. {\bf B#1} #2 }
\def\aph#1 #2 {Ann. Phys. {\bf #1} #2 }         
\def\jmp#1 #2 {J. Math. Phys. {\bf #1} #2 }
\def\jgp#1 #2 {J. Geom. Phys. {\bf #1} #2 }
\def\prd#1 #2 {Phys. Rev. {\bf D#1} #2 }
\def\prl#1 #2 {Phys. Rev. Lett. {\bf #1} #2 }
\def\rmp#1 #2 {Rev. Mod. Phys.  {\bf #1} #2 }
\def\zpc#1 {Z. Phys. {\bf #1C} }
\def\cmp#1 #2 {Commun. Math. Phys. {\bf #1} #2 }
\def\cqg#1 #2 {Class.Quant.Grav. {\bf #1} #2 }
\def\mpl#1 {Mod. Phys. Lett. {\bf A#1} }
\def\cpc#1 {Computer Phys. Commun. {\bf #1} }   % Belfast,cpc@v1.am.qub.ac.uk
\def\ijmp#1 {Int. J. Mod. Phys. {\bf A#1} }
\def\ijmpC#1 {Int. J. Mod. Phys. {\bf C#1} }

\def\BP{\begin{picture}} \def\EP{\end{picture}}         %% --> PICTURE macros 
\newcounter{TRefNX} \let\OLDcite=\cite  \makeatletter%       DRAFT MODE macros
\def\makeTRefs#1{\@for  \NewTRef:=#1\do{\global\makeTRef{\NewTRef}}}
\def\makeTRef#1{\ifundefined{TRef#1}\stepcounter{TRefNX}%
\expandafter\xdef\csname TRef#1\endcsname{\theTRefNX}\fi}\makeatother
\def\NEWcite#1{\makeTRefs{#1}\OLDcite{#1}}  

\ifundefined{draftmode} {} \else        \let\cite=\NEWcite
\newcount\HOUR  \HOUR=\the\time \divide\HOUR by 60      \multiply \HOUR by 60 
\newcount\MIN   \MIN=\the\time  \advance\MIN by -\HOUR  \divide\HOUR by 60
\ifnum  \MIN>9  \def\printTIME{{\it\the\HOUR\,:\,\the\MIN}}
        \else   \def\printTIME{{\it\the\HOUR\,:\,0\the\MIN}} \fi % \printTIME

\def\LLab#1{\BP(0,0)\unitlength=1mm\put(-12,.5){\makebox(0,0)[cr]{\small #1
        \rlap{$_{_{\makeatletter\csname TRef#1\endcsname\makeatother}}$}}}\EP}
\fi

\begin{document}

%\del

%\newfont{\XLbf}{cmbx10 scaled 2000}     \newfont{\XL}{cmr10 scaled 2000}
%\newfont{\XLbfmath}{cmmi10 scaled 2000}
%\newfont{\XHbf}{cmbx10 scaled 4800}     \newfont{\XH}{cmr10 scaled 4800}

\vspace*{-20mm}\begin{flushright}	\archive\\\preprint  \end{flushright}
\vspace*{-5mm}
\begin{center}	{\huge\bf	\title	}\end{center}
\begin{center} \vskip 3mm
        Maximilian KREUZER\fnote{*}{e-mail: maximilian.kreuzer@tuwien.ac.at}
\\[3mm]
        Institut f"ur Theoretische Physik, Technische Universit"at Wien\\
        Wiedner Hauptstra\3e 8--10, A-1040 Wien, AUSTRIA
\\[4mm]                       and
\\[4mm] Harald SKARKE\fnote{\#}{e-mail: skarke@zerbina.ph.utexas.edu}
\\[3mm] Theory Group, Department of Physics, University of Texas \\
        Austin, TX 78712, USA
        
\vfill\vfill	{\bf ABSTRACT } \end{center}\vspace{-2mm}    \abstract

\vfill\vfill \noindent \preprint\\[1pt] \finished \vfill
\thispagestyle{empty} \newpage
\pagestyle{plain} % \setcounter{page}{1} 

\newpage
\setcounter{page}{1}

%\enddel

\section{Introduction}

\subsection{Motivation}

When it was first realized \cite{CHSW} that manifolds of trivial 
canonical class play an important role in string compactifications,
only very few manifolds of this type were known. 
This situation changed with the first construction of large classes
of Calabi-Yau manifolds as hypersurfaces in weighted projective spaces
\cite{CLS}.
As these varieties are generically singular, it was not clear from a 
mathematical point of view what their Hodge numbers should be, 
but in string theory the corresponding numbers of generations and
antigenerations could be calculated by using orbifold techniques \cite{Vafa}.
These `physicists' Hodge numbers' showed a remarkable property that became 
known as mirror symmetry: to almost every manifold with a certain pair
of Hodge numbers there existed one or more other manifolds with the
Hodge numbers exchanged.
Soon an explicit construction applying to a subset of these spaces
%%%%	became available \cite{BH}. Only 
was found \cite{BH}. But	
when a complete classification of all such varieties was
available \cite{KlS, nms}
it became clear that mirror symmetry is not realized at the level 
of hypersurfaces in weighted projective spaces. 

Indeed, a far more natural setup for the
discussion of mirror symmetry is given in terms of toric geometry and
in particular by hypersurfaces in toric varieties that can by 
described by so-called reflexive polyhedra \cite{Bat}.
This construction not only implies manifest mirror symmetry (at the level of
Hodge number exchange), but also
explains how the weighted projective spaces have to be desingularised
(blown up) in order to allow smooth hypersurfaces with the Hodge
numbers that have been assigned to them through the orbifold construction.

It is often possible to obtain a Calabi-Yau manifold from another Calabi-Yau
manifold by first blowing down some divisors, thereby creating a singular
variety, and then resolving the singularity by changing the complex structure.
This raises the question of whether all Calabi-Yau manifolds might be
connected (directly or indirectly) by processes of this sort \cite{Reid},
which lead to important non-perturbative effects in string theory 
\cite{Str, GMS}.
It has been shown \cite{CGGK,ACJM}  that all Calabi-Yau threefolds that
are hypersurfaces in weighted projected spaces belong to a `web' of
this type.
This web could not be formed from hypersurfaces in weighted projective 
spaces alone, but required, once again, a generalization to toric
hypersurfaces.

It should be noted that almost all examples of manifolds of trivial 
canonical bundle occurring in the physics literature are hypersurfaces
(or, in a few cases, complete intersections) in toric varieties.
Therefore a classification of toric varieties that admit smooth 
three-dimensional hypersurfaces
of vanishing first Chern class is highly desirable.
Such a classification amounts to the classification of four-dimensional
reflexive polyhedra. 
While this classification problem is rather easy in two dimensions
(there are 16 well known reflexive polygons), only recently an
algorithm for approaching this problem in higher dimensions was found
\cite{crp,wtc}.
In the present work we fill in the last missing technical details of 
our algorithm and apply it to the classification of three dimensional 
reflexive polyhedra.
At first sight this might seem to be rather useless, since any hypersurface
resulting from this construction is going to be a K3 surface and all
K3 surfaces are known to be isomorphic with respect to their differential 
structures.
This is misleading, however, since in the context of
string dualities algebraic properties become important, and these 
algebraic properties are conveniently encoded in the structures of
the polyhedra.
When we consider F-theory or IIA duals to heterotic string
compactifications, we usually consider Calabi-Yau threefolds or
fourfolds that are K3 fibrations where the K3s themselves are
elliptically fibered.
Then the fibration structures manifest themselves as nestings of
the respective polyhedra \cite{CF,k3,fft}, and even the enhanced
gauge groups can be read off from the toric diagrams \cite{CF,egs,fst}.
Thus, the toric diagrams contain far more information than just 
which differential type of manifold we are dealing with.

In the remainder of the introduction we will give some definitions necessary
in the rest of the paper. 
In section 2 we give a rough outline of the strategy that we used for the 
classification and present our results.
In section 3 we explain our algorithm in more detail, 
starting with a summary of the
results of refs. \cite{crp,wtc} and then proceeding to more detailed
descriptions of various ideas that were relevant in the course of
implementing our algorithm.
% Finally, we present and discuss our results.

\subsection{Basic definitions}

A polytope in $\IR^n$ may be defined alternatively as the convex hull
of finitely many points or as an intersection of finitely many half spaces
that is bounded.
In the mathematics literature a polyhedron is also an intersection of 
finitely many half spaces, but not necessarily bounded \cite{Zie}.
We will, however, always mean `polytope' even when we write `polyhedron'.
More particularly, most of the polyhedra that we consider will be 
polytopes with $\ipo$ (the origin of $\IR^n$) in the interior. 
We will denote this property as the `interior point property' or `IP property'.
Given a polytope $\D$ in a vector space $M_\IR\simeq \IR^n$ with the IP 
property, we may define the dual (or polar) polytope 
$\D^*\subset N_\IR=M_\IR^*$ as
\beq 
\D^*=\{y\in N_{\IR}: ~~~\<y,x\>\ge -1 ~~~\forall x\in M_\IR\},  
\eeql{dual}
where $\<y,x\>$ is the duality pairing between $y\in N_{\IR}$ and $x\in M_\IR$.
Because of the convexity of $\D$, $(\D^*)^*=\D$.

Given a dual pair of polytopes such that $\D$ has $n_V$ vertices and $n_F$
facets (a facet being a codimension 1 face), the dual polytope has 
$n_V$ facets and $n_F$ vertices.
We may then define the vertex pairing matrix (VPM) $X$ as the $n_F\times n_V$
matrix whose entries are $X_{ij}=\<\bar V_i,V_j\>$, where $\bar V_i$ and
$V_j$ are the vertices of $\D^*$ and $\D$, respectively.
$X_{ij}$ will be $-1$ whenever $V_j$ lies on the i'th facet.
Note that $X$ is independent of the choice of a dual pair of bases in 
$N_{\IR}$ and $M_\IR$ but depends on the orderings of the vertices.

Given a lattice $M$, a lattice (or integer) polyhedron is a polyhedron on 
the real extension $M_{\IR}$ of $M$ whose vertices lie in $M$.
A lattice polyhedron $\D\subset M_{\IR}$ is called reflexive if its dual 
$\D^*\subset N_{\IR}$ is a lattice polyhedron w.r.t. the lattice $N$ dual
to $M$.
In this case the elements of the vertex pairing matrix $X$ are integer.
Note that, in turn, integer $X$ implies that there is a (finite number
of) lattice(s) with respect to which the polyhedron is reflexive. 
The coarsest such lattice is generated by the vertices of $\D$ and
is a sublattice of the finest lattice of this type, which is dual to 
the lattice generated by the vertices of $\D^ *$.

The lattice points of a reflexive polyhedron $\D$ encode the monomials
occurring in the description of the hypersurface in a variety whose
fan is determined by a triangulation of the dual polyhedron $\D^*$.
For details of what a fan is and how it determines a toric variety,
it is best to look up a standard textbook \cite{Ful, Oda}.

If a polyhedron $\D_1$ contains a polyhedron $\D_2$, then the definition
of duality implies $\D^*_1\subset \D^*_2$. 
Therefore the variety determined by the fan over $\D^*_1$ may be obtained
from the variety determined by the fan over $\D^*_2$ by blowing down
one or several divisors.
If we perform this blow-down while keeping the same monomials (those 
determined by $\D_2$), we obtain a generically singular hypersurface.
This hypersurface can be desingularised by varying the complex structure 
in such a way that we now allow monomials determined by $\D_1$.
Thus the classes of Calabi-Yau hypersurfaces determined by polyhedra $\D_1$ and
$\D_2$, respectively, can be said to be connected whenever $\D_1$ contains
$\D_2$ or vice versa.
More generally, if there is a chain of polyhedra $\D_i$ such that $\D_i$
and $\D_{i+1}$ are connected in the sense defined above, we call the
hypersurfaces corresponding to any two elements of the chain connected.

\section{Strategy and results}

Our approach to the classification of all reflexive polyhedra starts with
the construction of a set of maximal objects that contain all reflexive
polyhedra as subsets. Finding such a set in principle solves the classification
problem, but in practice the `trivial' second step of enumerating all 
reflexive subpolyhedra may be quite tricky or even impossible because of 
constraints of space and time. In section 3 we describe in
some detail which algorithms we used to complete the classification for the
3-dimensional case. But first we present a road map that shows how the 
pieces fit together.

Certain lattice polyhedra $\D$ can be described in the following 
simple way:
Take the
intersection of all positive half spaces $x_i\ge0$ with
the set of integer solutions to a linear equation $\sum q_i x_i = 1$ with 
positive rational coefficients $q_i>0$,
and define $\D$ to be the convex hull of these points.
% (this can be interpreted as the Newton polyhedron of a quasi-homogeneous 
% polynomial of degree $d$).
If $\sum q_i=1$ then $\D$ has at most one 
interior lattice point,
namely the point $\ip$ with all coordinates $x_i=1$.
%  is contained in $\D$ and it is the only 
% candidate for an interior point (IP).
% There are two crucial facts that make our construction go \cite{crp,wtc}: \\
Our approach is based upon two crucial facts \cite{crp,wtc}: \\
{\it1.} In any dimension $n$ there is only a finite number of (single) weight 
	systems $(q_i)$ with $\sum q_i=1$ 
	such that \ip\ is in the interior of the corresponding polyhedra.
	%have an interior point. 
	(By definition one 
	interior lattice point
	is necessary for 
	reflexivity; in $n\le4$ dimensions it is also sufficient for
	polyhedra of this type \cite{wtc}.)\\
{\it2.} Each reflexive polyhedron is contained in an object that is slightly 
	more general: We may have to embed it into $\IZ^{k}$ with codimension
	$k-n>1$ using sets of solutions to $k-n$ equations of the type 
	$\sum q_j^{(i)} x_j = 1$. There is a finite number of possible types 
	of such combined weight systems, which consist of $k-n$ single weight 
	systems which are extended by zeros (see section 3.1 or \cite{crp}).\\
The last entry in table 1, for example, 
corresponds to a cube embedded in $\IR^6$ by 
\beq 
x_1+x_2=2,~~~ x_3+x_4=2,~~~ x_5+x_6=2~~~\hbox{ and }~~~x_i\ge 0~~\forall i.
\eeq
In the three examples of table 1 with bi-degrees 
$(d^{(1)},d^{(2)})$ equal to (3,3), (3,4) and (4,4), respectively,
the coordinate $x_1$ enters both equations.

\begin{figure}[htb]
\begin{center}
\begin{tabular}{||c||c|c|c|c|c|c|c|c|c||c|c|c|c|c||c||}\hline\hline
$d^{(i)}$  &	4 & 5 & 6 & 6 & 7 & 8 & 9 &10 &12 
	&	3 2 & 4 2 & 3 3 & 3 4 & 4 4 		& 2 2 2 \\\hline
$w_1^{(i)}$&	1 & 1 & 1 & 1 & 1 & 1 & 1 & 1 & 1
	&	1 0 & 1 0 & 1 1 & 1 2 & 2 2 		& 1 0 0	\\
$w_2^{(i)}$&	1 & 1 & 1 & 1 & 1 & 1 & 1 & 1 & 1 
	&	1 0 & 1 0 & 1 0 & 1 0 & 1 0 		& 1 0 0	\\
$w_3^{(i)}$&	1 & 1 & 2 & 1 & 2 & 2 & 3 & 3 & 4 
	&	1 0 & 2 0 & 1 0 & 1 0 & 1 0 		& 0 1 0	\\
$w_4^{(i)}$&	1 & 2 & 2 & 3 & 3 & 4 & 4 & 5 & 6 
	&	0 1 & 0 1 & 0 1 & 0 1 & 0 1 		& 0 1 0	\\
$w_5^{(i)}$ &&&&& &&&&&	  0 1 & 0 1 & 0 1 & 0 1 & 0 1	& 0 0 1 \\
$w_6^{(i)}$ &&&&& &&&&&   & & & & &			  0 0 1 \\
\hline\hline	Points	& 35,\,19&34&30&39&31&35&33&36&39 & 30&27&30&31&35 	& 27\\
\hline\hline\end{tabular}
\\[5mm]
{{\bf Table 1: }The 9 single and the 6 combined weight systems defining
	the polytopes \\ containing all others, and the respective numbers 
        of points.~~
}
\end{center}
\end{figure}

According to \cite{wtc,web}, there are 58 single and 21 combined
weight systems relevant to our classification scheme. 
Our first new result is that 
these numbers may still be reduced: It turns out that
all 3-dimensional reflexive
polyhedra are contained in the 15 polyhedra that are defined by the weight
systems in table 1 (cf. section 3.2). 
There is, however, a subtle point: This statement is true only if we also 
admit %convex subsets that live on 
sublattices
of the lattice that is defined by 
integer %ambient space coordinates 
$x_i$. %\in\IZ$.
Indeed, it turns out that there is 
one %(maximal) 
polytope 
that we would miss if we ignored sublattices:
It is a $\IZ_2$ quotient with 19 
lattice points of 
the simplex
\beq \{(x_1,x_2,x_3): x_i\ge-1 \and x_1+x_2+x_3\le1	\}\eeq
with 35  lattice
points that %corresponds to 
 is defined by 
the single weight system with degree 4 (the coordinates
have been shifted by 1 such that the interior lattice point is at the origin).
To obtain the $16{th}$ 
 polytope that is not a subpolytope of any other, we can 
restrict our lattice, for example, to
%apply, for example, the $\IZ_2$ projection 
$x_1+x_2\equiv 0 \mod 2 $, which keeps 19 of the 35 points, % of the quartic,
including all vertices and \ipo. 
%Geometrically this means that 
 In other words,
we take out every other lattice plane parallel 
to a fixed set of two non-intersecting edges. Because of the full permutation 
symmetry of the 4 vertices there are 3 different choices of such a plane, which
lead to the same polytope up to lattice automorphisms. 
In terms of toric geometry, we have the following interpretation:
The simplex with 35 lattice points is the Newton polytope of the 
quartic hypersurface in $\IP^3$.
The $\IZ_2$ quotient of lattices corresponds to a $\IZ_2$ quotient of toric
varieties.
$\IP^3/\IZ_2$ has singularities at the fixed lines ($\IP^1$s) $z_1=z_2=0$ and
$z_3=z_4=0$ of the $\IZ_2$ action (the $z_i$ being the homogeneous 
coordinates of the $\IP^3$).
These singular lines must be blown up to obtain a smooth
toric variety in which we have a K3 hypersurface whose Newton polytope is 
the simplex with 19 lattice points.

The fact that every polyhedron $\D$ is contained in at least one of the 16 
objects we just discussed implies that the dual $\D^*$ contains one
of the duals of these 16 polytopes.
Therefore, the fan of any toric ambient variety determined by a maximal
triangulation of a reflexive polyhedron is a refinement of one of the 
corresponding 16 fans.
In other words, any such toric ambient variety is given by the blow-up 
of one of the following 16 spaces  (cf. table 1):\\
-- $\IP^3$, \\
-- $\IP^3/\IZ_2$, \\
-- 8 different weighted projective spaces $\IP^2_{(q_1,q_2,q_3)}$,\\
-- $\IP^2\times\IP^1$,  \\
-- $\IP^2_{(1,1,2)}\times\IP^1$, \\
-- 3 further double weighted spaces, and\\
-- $\IP^1\times\IP^1\times\IP^1$. \\
Each of the three spaces with `overlapping weights' allows two distinct
bundle structures: The first one can be interpreted as a $\IP^2$ bundle
in two distinct ways, the second one as a $\IP^2$ bundle or a 
$\IP^2_{(1,1,2)}$ bundle, and the third one can be interpreted as a 
$\IP^2_{(1,1,2)}$ bundle in two distinct ways.
In each case the base space is $\IP^1$.

In order to enumerate all 3-dimensional reflexive polytopes we thus 
had to construct all lattice 
subpolytopes $\D$ of the 15 objects defined 
by the weight systems in table 1 such that $\D$ is 
reflexive on some lattice. 
\del
This means that, in a first step, we are actually classifying integer 
vertex pairing matrices (VPMs), from which we can then, in a second step,
reconstruct the reflexive polytopes that define them. It is, nevertheless,
useful to store the polytopes rather than the VPMs because this allows us
to check for each polyhedron if we already searched the convex subsets of 
another object that only differs by a lattice automorphism and to avoid the 
resulting redundancy. 
Starting with the 15 weight systems we thus 
\enddel
We first found  6202 
inequivalent subpolytopes with integral VPM
(polytopes that are reflexive on some lattice; cf. section 1.2),
4318 of which are reflexive
on the original lattice.
Then we computed the resulting 4075 inequivalent VPMs
(to obtain these numbers we defined and computed normal forms of the 
respective objects and wrote them into a sorted list; cf. section 3.4).
Going over all allowed sublattices for all integer VPMs 
(for details see section 3.5)
we eventually constructed all 4319 reflexive polytopes.
The complete list is accessible via internet.%
\footnote{	It can be found at 
{\tt
	http://tph16.tuwien.ac.at/\~{}kreuzer/CY.html}
} 
Some statistics is compiled in table 2.

\begin{figure}[htb]
\begin{center}
\begin{tabular}{||c||c|c|c|c|c|c|c|c|c|c|c||}\hline\hline
Points		& 5 & 6 & 7 & 8 & 9 & 10 & 11 & 12 & 13 & 14 & 15\\\hline
Multiplicity	& 1 & 7 &23 &54 &135&207 &314 &373 &416 &413 &413\\\hline\hline
\hline
Points		& 16 & 17 & 18 & 19 & 20 & 21 & 22 & 23 & 24 & 25 & 26\\\hline
Multiplicity	&348 &334 &274 &234 &179 &151 &117 & 87 & 66 & 40 & 42\\\hline
\hline
\hline Points	& 27 & 28 & 29 & 30 & 31 & 32 & 33 & 34 & 35 & 36 & 39\\\hline
Multiplicity	& 27 & 18 &  8 & 13 &  9 &  4 &  2 &  2 &  5 &  1 & 2 \\
\hline\hline\end{tabular}
\\[5mm]
{{\bf Table 2: }Multiplicities of point numbers for the 4319 reflexive
	polytopes.
}
\end{center}
\end{figure}

There are several reasons why we decided to store 6202 polytopes rather 
than 4075 VPMs in the first step of the enumeration process.
The fact that polyhedra require less disc space might become important 
in the context of four dimensional polyhedra. 
Besides, having the polyhedra explicitly 
allows us
to check for each polyhedron if we already searched the convex subsets of 
another object that only differs by a lattice automorphism and to avoid the 
resulting redundancy. 
%Sticking to a fixed lattice in the first step of our construction also had 
The most important advantage is
that we could easily check for connectedness: For each new
weight system we 
checked explicitly that at least one of its subpolyhedra had been found
\del
incremented some counter whenever we came across a
reflexive polytope that already was found for a different weight system
\enddel
before. Connectedness of the original list of 4318 polytopes follows from the 
fact that this %`hit count' was always positive.
was always the case.
All 679 reflexive proper subpolytopes of the exceptional polytope that we 
only found on a sublattice already show up as subpolyhedra of the 15 maximal 
objects that live on the original lattices. This establishes connectedness of 
the complete set of 4319 reflexive polyhedra in 3 dimensions.

An important check for % completeness of our construction 
the correct implementation of our classification algorithm
is mirror symmetry, 
i.e. that we obtain for each 
of the 4319 polyhedra the dual one in the sense of (\ref{dual}).
For convenience, we actually %We explicitly
checked a slightly weaker statement, namely that we got
the dual (i.e., transposed) for each of the 4075 distinct VPMs.
The fact, however, that
we recovered all 4318 previously found reflexive polyhedra 
(and, in addition, found the new one) from these VPMs, 
also provides a very stringent test for the last step of our 
construction.

In our first complete calculation we stored all 665598 different
subpolytopes (on the original lattices) that have an interior point. This 
took about 2 hours of CPU time and required more than 60 MB of memory. 
If we only remember the subpolytopes with integer VPM (and thus risk to 
reanalyze the same non-reflexive objects again and again) computation time 
triples, but in return we only need less than 1 MB of RAM. 

Thinking about the 4-dimensional case, where we have to deal with 
at least
308 weight systems and with respective point numbers between 47 and 680,
this means, of course, that our 
program has to be improved drastically. From the above numbers it is
clear that we will have to avoid the construction of all IP subpolytopes.
In an improved version we terminated the iteration whenever we could verify
that all subpolytopes in the present branch would have a facet with distance
larger than one from the IP. This reduced the computation time to below 1
minute. With this improved program we already
produced about 20 million reflexive 4-dimensional polytopes and the complete 
number may well be as large as $10^9$, which would surpass the computing 
resources that presently are at our disposal. Nevertheless, even with an
incomplete list, it will be interesting how the spectrum of 
Hodge
numbers of
the corresponding Calabi--Yau manifolds changes as compared to the relatively
few examples that have been known so far.

\begin{figure}[htb]
\begin{center}
\begin{tabular}{||c||c|c|c|c|c|c|c|c|c|c||}\hline\hline
Picard number   & 1 & 2 & 3 & 4 & 5 & 6 & 7 & 8 & 9 & 10 \\\hline
Multiplicity    & 2 & 9 &25 &58 &101&165&254&372&489&574 \\\hline
                                                         \hline\hline
Picard number   & 11 & 12 & 13 & 14 & 15& 16& 17& 18& 19&\\\hline
Multiplicity    &578 &521 &451 &350 &204&112& 40& 12& 2 &\\\hline
\hline\end{tabular}
\\[5mm]
{{\bf Table 3: }Multiplicities of Picard numbers for the 4319 reflexive
        polytopes.
}
\end{center}
\end{figure}
Returning to the case of K3 surfaces, we mention that we also 
calculated the Picard number for each of our 4319 models,
using the formula \cite{Bat}
\beq \hbox{Pic}=l(\D^*)-4-\sum_{{\rm facets}~\th^*~{\rm of}~\D^*}l^*(\th^*)+
     \sum_{{\rm edges}~\th^*~{\rm of}~\D^* }l^*(\th^*)l^*(\th),
\eeq
where $l$ denotes the number of integer points of a 
polyhedron and $l^*$ denotes the number of interior integer points of a 
facet or an edge.
Contrary to the case of higher dimensional Calabi-Yau manifolds,
this number is not the same as the Hodge number $h_{11}$, which is always
20 for K3 surfaces. Instead,
\beq 
h_{11}=20=\hbox{Pic}+l(\D)-4-\sum_{{\rm facets}~\th~{\rm of}~\D}l^*(\th),
\eeq
which is another useful check on our programs.
Mirror symmetry for K3 surfaces is usually interpreted in terms of 
families of lattice polarized K3 surfaces 
(see, e.g., \cite{Dol} or \cite{Arev}).
In this context the Picard number of a generic element of a family and the 
Picard number of a generic element of the mirror family add up to 20.
The fact that the Picard numbers for toric mirror families add up to
$20+\sum l^*(\th^*)l^*(\th)$ indicates that our toric models
occupy rather special loci in the total moduli spaces.

Let us end this section with briefly discussing a few of the most interesting
objects in our list.
As table 3 indicates, there are precisely two mirror pairs with Picard numbers
1 and 19, respectively.
One of them is the quartic hypersurface in $\IP^3$ with Picard number 1,
together with its mirror of Picard number 19, which is also the model 
whose Newton polytope is the only reflexive polytope with only 5 
lattice points.
This model corresponds to a blow-up of a $\IZ^4\times\IZ^4$ orbifold
of $\IP^3$.
The blow-up of six fixed lines $z_i=z_j$ by three divisors each yields
18 exceptional divisors leading to the total Picard number of 19.
The other mirror pair with Picard numbers 1 and 19 consists of the hypersurface
in $\IP^3_{(1,1,1,3)}$ of degree 6 and an orbifold of the same model, with
Newton polyhedra with 39 and 6 points, respectively.
The other polyhedron with the maximal number of 39 points is the Newton 
polytope of the hypersurface of degree 12 in $\IP^3_{(1,1,4,6)}$.
This model leads to the description of elliptically fibered K3 surfaces
that is commonly used in F-theory applications \cite{F,MVI,MVII},
with the elliptic fiber embedded in a $\IP^2_{(1,2,3)}$ by a Weierstrass
equation.
The mirror family of this class of models can be obtained by forcing two
$E_8$ singularities into the Weierstrass model and blowing them up.
The resulting hypersurface allows also a different fibration structure
which can develop an $SO(32)$ singularity; thereby this model is able
to describe the F-theory duals of both the $E_8\times E_8$ and the $SO(32)$
heterotic strings with unbroken gauge groups in 8 dimensions \cite{fst}.

\section{The algorithm}

\subsection{General outline and existing results}

The starting point of our algorithm is the introduction of the concept
of a minimal polyhedron \cite{crp}.
Consider a polyhedron in $\IR^n$ with the interior point (IP) property.
We call this polyhedron minimal if there is no strict subset 
$\{V_i,i\in I\},I\subsetneq \{1,\ldots,k\}$
of the set $\{V_1,\ldots,V_k\}$ of vertices such that the convex hull
of $\{V_i,i\in I\}$ has $\ipo$ in its interior.
It could be shown that these objects allow a classification according to the 
types of linear relations between its vertices.
In particular, in two dimensions only two types are possible:
The triangle $V_1V_2V_3$ with 
\beq q_1V_1+q_2V_2+q_3V_3=0 \qquad\hbox{ where }\qquad 
	0<q_i<1,\qquad q_1+q_2+q_3=1 
\eeq 
and the parallelogram $V_1V_2V_1'V_2'$ with 
\beq q_1V_1+q_2V_2=0,\qquad q_1'V_1'+q_2'V_2'=0 \qquad\hbox{ where }\qquad   
     0<q_i,q_i'<1, \qquad      q_1+q_2=q_1'+q_2'=1.  \eeq
In a shorthand notation, this result may be summarized as $\{$3; 2+2$\}$.
In the same notation the result for three dimensions can be summarized
as $\{$4; 3+2, $\cont{\hbox{3+3}}$; 2+2+2$\}$ where the underlining symbol 
indicates a vertex occurring in both linear relations. 
Here this means that our polyhedron is the
convex hull of $V_1,V_2,V_3,V_2',V_3'$ and that there are relations
$q_1V_1+q_2V_2+q_3V_3=0$ and $q_1'V_1+q_2'V_2'+q_3'V_3'=0$.
A minimal polyhedron $\nabla_{\rm min}\subset \IR^n$ is then specified (up to 
linear transformations) by its structure and
by the weights $q_i$ involved in this construction.
We will call (combinations of) sets of $q_i>0$ with $\sum q_i=1$
(combined) weight systems.
Notice that everything said so far applies to polyhedra in $\IR^n$
and real $q_i$. 
In the context of lattice polyhedra (which we will consider henceforth)
the $q_i$ are rational.

Clearly any polyhedron with the IP property allows at
least one (possibly trivial) subset of vertices whose convex hull is
a minimal polyhedron.
Applying this to the dual $\D^*$ of a reflexive polyhedron, we find that 
there exists a minimal integer (not necessarily reflexive) polyhedron 
$\nabla_{\rm min}\subseteq\D^*$, implying $\D\subseteq\nabla_{\rm min}^*$.
The fact that $\D$ is a lattice polyhedron leads to the stronger restriction
\beq \D\subseteq\Dm:= \hbox{ConvexHull}(\nabla_{\rm min}^* \cap M).  \eeq
Given a minimal polyhedron $\nabla_{\rm min}\subset \IR^n$ we still have to
specify a choice of lattice $N\subset N_\IR\simeq \IR^n$.
The coarsest possible such lattice $N_{\rm coarsest}$ is the lattice
generated by the vertices of $\nabla_{\rm min}$.
Its dual is the finest $M$ lattice $M_{\rm finest}$. 
Any other $M$ lattice compatible with integrality of $\nabla_{\rm min}$ 
is a sublattice of $M_{\rm finest}$.

For pairs of reflexive polyhedra clearly only minimal polyhedra and
therefore (combined) weight systems such that $\Dm$ (w.r.t. $M_{\rm finest}$)
has the IP property are relevant.
In such a case we also say that a (combined) weight system has the IP
property.
As a side remark we mention that this definition implies
reflexivity of $\Dm$ for lattice dimensions $n\le4$ \cite{wtc}.
It is easy to see that a combined weight system can have the IP property
only if each of its weight systems by itself has it.
The weight systems with up to 5 weights with the IP property
were classified in \cite{wtc}.
There is one such system $(1/2,1/2)$ with two weights; with three weights there
are the three systems $(1/3,1/3,1/3)$, $(1/2,1/4,1/4)$, $(1/2,1/3,1/6)$;
there are 95 systems with four and 184,026 systems with five weights.

Given all simple weight systems, finding the combined weight systems is
an easy combinatorial task \cite{web}. From 
what we have discussed up to now it is clear that any reflexive polyhedron
is a subpolyhedron of a maximal polyhedron $\Dm$ w.r.t. some (combined)
weight system $(q)$ that has the IP property, on some 
lattice $M$ that is $M_{\rm finest}$ or a sublattice thereof.
Thus the task of classifying reflexive polyhedra is reduced to the task of
classifying all reflexive subpolyhedra $\D\subset M\subseteq M_{\rm finest}$ 
of all maximal polyhedra $\Dm\subset M_{\rm finest}$.
Before we describe how to find all subpolyhedra of a given polyhedron
efficiently, we will show that the set of (combined) weight
systems relevant for the classification scheme may still be reduced.

\subsection{New definitions of minimality of polyhedra}

Remember that our definition of a minimal polyhedron meant that no subset
of the set of vertices of $\nabla_{\rm min}$ should define a polyhedron with 
the IP property 
(we might have called such polyhedra vertex-minimal).
If we work not just in $\IR^n$ but with a lattice, we may similarly
define a polyhedron $\nabla_{\rm lpm}$ to be lp-minimal (lp standing for 
`lattice point') if no subset of the set 
of lattice points of $\nabla_{\rm lpm}$ defines a polyhedron with the interior 
point property. 
A (combined) weight system will be called lp-minimal if the corresponding
$\nabla_{\rm min}$ on $N_{\rm coarsest}$ is lp-minimal.
Clearly a polyhedron that is not lp-minimal will contain an lp-minimal
polyhedron as a proper subset; therefore only lp-minimal polyhedra will
play a role in our classification scheme.

Note, however, that even being lp-minimal does not guarantee that 
$\Dm$ is not a subpolyhedron of any other reflexive polyhedron:
$\Dm^*$ might contain not only $\nabla_{\rm lpm}$, but also a different
lp-minimal polyhedron $\tilde\nabla_{\rm lpm}$ such that 
$\Dm\subseteq\tilde\Dm$.
This cannot happen, however, if 
it is impossible to omit any of the vertices of $\nabla_{\rm min}$ from 
$\Dm^*\cap N$ without violating the IP property.
In that case we
call the corresponding (combined) weight systems very minimal.
Clearly very minimal implies lp-minimal.

As an example of a polyhedron that is lp-minimal but not very minimal 
consider $\nabla_{\rm min}$ defined by the single weight system 
(1,2,3,5)/11.
Here we may represent the vertices of $\nabla_{\rm min}$ by
\beq V_1=(1, 0, 0),\; V_2=(0, 1, 0),\; V_3=(0, 0, 1) \;\hbox{and} \;
     V_4=(-2,-3,-5), \eeq
the only other lattice point 
being \ipo.
As $\nabla_{\rm min}^*$ is not a lattice polyhedron, $\Dm$ is smaller than
$\nabla_{\rm min}^*$ and so $\Dm^*$ is larger than $\nabla_{\rm min}$.
More precisely, $\Dm^*$ has the three additional vertices 
\beq V_5=(-2,-3,-6),\; V_6=(0,-1,-1)\; \hbox{and}\; V_7=(-1,-1,-2) \eeq
and besides contains 6 further lattice points.
Thus $\nabla_{\rm min}^{(1,2,3,5)/11}$ contains 
$\nabla_{\rm min}^{(1,2,3,6)/12}$
(the convex hull of $V_1, V_2, V_3, V_5$)
and $\nabla_{\rm min}^{(1,1,1,2)/5}$ (the convex hull of $V_1, V_2, V_3, V_7$).
In addition it contains some other minimal polyhedra corresponding
to other lattice points.
Therefore $\Dm^{(1,2,3,5)/11}$ is contained in other maximal polyhedra and
does not play a role in the classification algorithm.

An example of a very minimal weight system is (1,1,3,4)/9.
The vertices of $\nabla_{\rm min}$ can be represented as
\beq V_1=(1, 0, 0),\; V_2=(0, 1, 0),\; V_3=(0, 0, 1) \;\hbox{and} \;
     V_4=(-1,-3,-4), \eeq
the only other lattice point again being \ipo.
$\Dm^*$ has the additional points 
\beq V_5=(0,-2,-3),\; P_6=(0,-1,-2),\; P_7=(0,0,-1) \;\hbox{and} \;  
     P_8=(0,0,-1) \eeq
($V_5$ is the only vertex among them).
Dropping any of the vertices $V_1,\ldots,V_4$ from  $\Dm^*$ results in
loss of the IP property, as is easily checked.
Thus there cannot be a weight system $(q)\ne (1,1,3,4)/9$ such that 
$(\Dm^{(q)})^*\subseteq(\Dm^{(1,1,3,4)/9})^*$ or, conversely, that 
$\Dm^{(1,1,3,4)/9}\subseteq\Dm^{(q)}$.

Let us summarize: 
By the analysis given in this subsection,
every reflexive polyhedron is contained in the dual of
an lp-minimal polyhedron, and the duals of very minimal polyhedra are not 
contained in the duals of any other minimal polyhedra.

Our computer programs led to the following further statements: For 
$n=3$, there are 15 very minimal weight systems and 4 further weight systems
that are lp-minimal without being very minimal.
The latter weight systems lead to polytopes $\Dm$, however, that are contained
in the $\Dm$ coming from the 15 very minimal weight systems.

\del	In $d\le4$ dimensions the very minimal polyhedra automatically define
	reflexive objects, which comprise the set of maximal reflexive
	polyhedra, obviously containing all others. 
	In any case it can be checked a posteriori if all duals of lp-minimal 
	polyhedra were found among the subpolytopes of the duals of very minimal
	ones. ~~~
\enddel

\subsection{Finding subpolyhedra}

Finding all subpolyhedra of a given polyhedron without introducing 
excessive redundancy is a non-trivial task. 
Postponing the problem of identifying polyhedra that are related by
lattice automorphisms to the next section, the first step is to
construct all convex subsets of lattice points. This can be 
 achieved 
by first generalizing the problem: Consider the 
task of finding all subpolyhedra of a given polyhedron $\D$ that contain
a specific subset of the vertices of $\D$ (let us call these allowed
subpolyhedra).
We may think of specifying this subset by attaching labels $k$ (for `keep')
to these vertices.
Finding the corresponding subpolyhedra of $\D$ can be done recursively: 
Unless $\D$ is empty or all of the vertices of $\D$ carry $k$ labels, we 
may choose a vertex $V$ that does not carry a $k$ label.
Then the allowed subpolyhedra of $\D$ will be the allowed subpolyhedra
of $\D\setminus\{V\}$ and the allowed subpolyhedra of $\D$ with an 
extra $k$ label attached to $V$.
Clearly the recursion terminates only when $\D$ is empty (or, in our 
case we let it terminate whenever it ceases to have the interior point
property) or when all of its vertices carry $k$ labels.

Our original problem can thus be solved with the following recursive \
algorithm:\\
(1) Start with $\D_{\rm max}$ without labels\\
(2) At a specific step in the recursion we have a polyhedron $\D$,
some of whose vertices carry $k$ labels.
If $\D$ doesn't have the IP property, we don't continue with it.
Otherwise, if all vertices of $\D$ have $k$ labels, 
we can add $\D$ to the list of subpolyhedra of $\D_{\rm max}$.
Otherwise, pick a specific vertex $V$ without a label.
Do step (2) with the label of $V$ changed to $k$ and with 
$\D'=\D\setminus\{V\}$.

As a vertex of a polyhedron is also a vertex of any subpolyhedron to
which it belongs, a point that carries a $k$ label is always guaranteed
to be a vertex.

The application of this algorithm to the Newton polygon of $\IP_{(1,2,3)}$
is shown pictorially in figure 1. 
Note how every subpolyhedron occurs precisely once as an endpoint of the 
recursive tree.
\begin{figure}[htb]
\epsfxsize=4in
\hfil\epsfbox{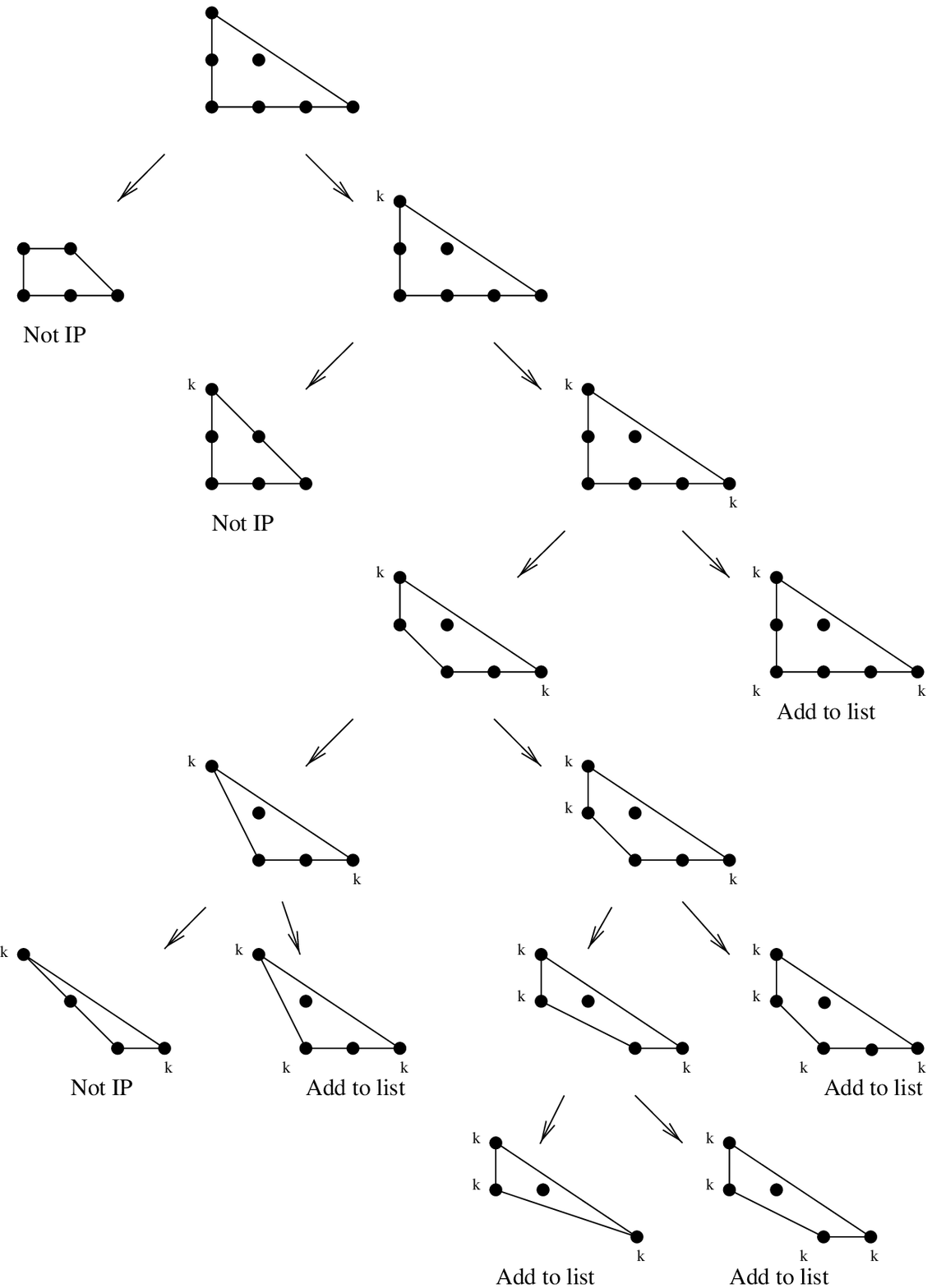}\hfil
\caption{\it Applying the classification algorithm to $\D(\IP_{(1,2,3)})$}
\label{class}
\end{figure}

\subsection{Normal forms of polyhedra}

In a classification scheme like the present one that produces the 
complete set with large redundancy it is useful to define 
normal forms of the relevant objects. A normal form 
%automatically implies 
allows us to define
a total ordering (for example lexicographic) so that we can
efficiently check for new entries by searching a sorted list with 
bisection. 

To describe our polyhedra we start with the matrix of coordinates 
of the vertices. This matrix is determined 
only up to an $S_{n_V}\times GL(d,\IZ)$ symmetry, where $S_{n_V}$ is
the group of permutations of the vertices and $GL(d,\IZ)$ is
the group of coordinate transformations of a $d$-dimensional lattice.
% In order to be able to compare two polyhedra, it is useful to lift this
% degeneracy by defining some sort of `normal form' of the list of vertices.
To lift this redundancy we first define a normal form for vertex pairing 
matrices using permutations of lines and columns to obtain the 
(lexicographically) maximal matrix. This removes  
% This allows us to remove most of 
the $S_{n_V}$ degeneracy, except for the subgroup that corresponds to the 
symmetry of the polyhedron on the coarsest lattice, if we demand
that the ordering of the vertices should be the same as in the normal 
form of the vertex pairing matrix.
Given a particular ordering of the vertices we then use the $GL(d,\IZ)$ 
freedom to make the matrix of vertices upper diagonal with positive
elements along the diagonal and minimal non-negative elements above.
In the cases where the normal form for vertex pairing matrices has not
determined the ordering of vertices unambiguously, we do this for
every allowed ordering and choose the lexicographically smallest one to
be our normal form. 
Invariances of this normal form correspond to symmetries of the polyhedron. 

To make sure that we cannot miss any reflexive polyhedron in our 
classification scheme we have to keep record
of any polyhedron that is reflexive either on the lattice on which it
was found or on any sublattice.
A necessary and sufficient condition for a polyhedron to be reflexive on 
some lattice is just integrality of the \vpm: 
If $\D$ is reflexive, then its \vpm\ will obviously be integer, and if
the \vpm\ is integer, $\D$ will be reflexive on the lattice generated
by the vertices of $\D$.
So the minimal set of polyhedra to be stored is the set of those 
with integer \vpm.
For saving calculation time, however, it makes sense to keep record of 
non-reflexive polyhedra as well, thereby increasing the required 
storage capacity.
For the present case of classifying three-dimensional reflexive 
polyhedra it doesn't really matter if one chooses to keep record of 
non-reflexive polyhedra.
It seems, however, that for four-dimensional polyhedra memory seems
to be a greater problem than time, so it probably never makes sense
to keep record of polyhedra that are not reflexive on any lattice.

\subsection{Sublattices}

Given an integer vertex pairing matrix, there is always a coarsest 
lattice on which the polyhedron $\D$ is reflexive (the lattice generated 
by the vertices of $\D$) and a finest lattice (the lattice dual to the one
generated by the vertices of $\D^*$).
The quotient of these lattices is a finite Abelian group.
This group can be represented by expressing the generators
of the finer lattice in terms of the generators of the coarser lattice.
The corresponding vectors in $\IQ^n$, taken modulo $\IZ^n$, are the
generators of the quotient group.

A vertex pairing matrix $X$ is an $n_F\times n_V$ matrix, $n_F$ and $n_V$
being the numbers of facets (dual vertices) and vertices, respectively.
$X$ can be decomposed $X=W\cdot D\cdot U$ where $W$ is $n_F\times d$,
$D$ is a $d\times d $ diagonal matrix and $U$ is $d\times n_V$.
Before giving a geometrical interpretation, let us see how
this decomposition can be achieved algorithmically:
By recombining the lines and columns of $X$ in the style of Gauss's
algorithm for solving systems of linear equations, we can
turn $X$ into an $n_F\times n_V$ matrix $\tilde D$ with non-vanishing 
elements only along the diagonal.
But recombining lines just corresponds to left multiplication with 
some $GL(\IZ)$ matrix, whereas recombining columns corresponds to right
multiplication with some $GL(\IZ)$ matrix.
Keeping track of the inverses of these matrices, we successively create
decompositions $X=\tilde W^{(n)}\cdot \tilde D^{(n)}\cdot \tilde U^{(n)}$ 
(with $\tilde W^{(0)}=1$, $\tilde D^{(0)}=X$ and $\tilde U^{(0)}=1$).
Let us denote the matrices resulting from the last step by 
$\tilde W$, $\tilde D$ and $\tilde U$.
$\tilde W$ and $\tilde U$ being regular matrices and the rank of $X$ being
$d$, it is clear that $\tilde D$ has only $d$ non-vanishing elements
which can be taken to be the first $d$ diagonal elements.
Then we can choose $W$ to consist of the first $d$ columns of $\tilde W$,
$U$ to consist of the first $d$ lines of $\tilde U$ and $D$ to be the
upper left $d\times d$ block of $\tilde D$.

The interpretation of $U$ and $W$ is as follows: 
We may view the columns of $X$ as the coordinates of $\D$ (on $M_{\rm finest}$)
in an auxiliary $n_F$-dimensional space carrying an $n_F$-dimensional 
lattice in which $\D$ is embedded. 
The $n_F\times n_F$ matrix $\tilde W$ effects a change of
coordinates in the $n_F$-dimensional lattice so that $\D$ now lies in
the lattice spanned by the first $d$ coordinates.
Thus we can interpret the columns of $D\cdot U$ as the vertices of $\D$ on 
$M_{\rm finest}$.
Similarly, the lines of $W\cdot D$ are coordinates of the vertices of 
$\D^*$ on $N_{\rm finest}$, whereas $U$ and $W$ are the corresponding 
coordinates on the coarsest possible lattices.

More explicitly, denoting the generators of $M_{\rm coarsest}$ by $\1 E_i$ 
and the generators of $M_{\rm finest}$ by $\1e_i$, we have 
$\1 E_i=\1e_jD_{ji}$.
An intermediate lattice will have generators $\1\ce_i=\1e_jT_{ji}$
such that the $\1 E_i$ can be expressed in terms of the $\1\ce_j$,
amounting to 
\beq \1 E_i=\1\ce_jS_{ji}=\1e_kT_{kj}S_{ji}     \eeq
with some integer matrix $S$.
This results in the condition $D_{ki}=T_{kj}S_{ji}$.
In order to get rid of the redundancy coming from the fact that the 
intermediate lattices can be described by different sets of generators,
one may proceed in the following way: 
$\1\ce_1$ may be chosen as a multiple of $\1e_1$ (i.e., 
$\1\ce_1=\1e_1T_{11}$).
Then we choose $\1\ce_2$ as a vector in the $\1e_1$-$\1e_2$-plane 
(i.e., $\1\ce_1=\1e_1T_{12}+\1e_2T_{22}$) subject to the
condition that the lattice generated by $\1\ce_1$ and $\1\ce_2$ should be 
a sublattice of the one generated by $\1E_1$ and $\1E_2$, which is 
equivalent to the possibility of solving $T_{kj}S_{ji}=D_{ki}$ for integer 
matrix elements of $S$.
We may avoid the ambiguity arising by the possibility of adding a multiple of
$\ce_1$ to $\ce_2$ by demanding $0\le T_{12}<T_{11}$.
We can choose the elements of $T$ column by column (in rising order).
For each particular column $i$ we first pick $T_{ii}$ such that
it divides $D_{ii}$; then $S_{ii}=D_{ii}/T_{ii}$.
Then we pick the $T_{ji}$ with $j$ decreasing from $i-1$ to $1$.
At each step the $j$'th line of $T\cdot S =D$,
\beq T_{ji}S_{ii}+\sum_{j<k<i}T_{jk}S_{ki}+T_{jj}S_{ji}=0, \eeq
must be solved for the unknown $T_{ji}$ and $S_{ji}$ with the
extra condition $0\le T_{ji}<T_{ii}$ ensuring that we get only one
representative of each equivalence class of bases.
Proceeding in this way, we create all inequivalent
upper triangular matrices $T$ and $S$
such that the coordinates of the vertices of $\D^*$ on the intermediate 
lattice given by the $\1\ce_i$ are the columns of $S\cdot U$ and the 
vertices of $\D^*$ on the corresponding dual lattice are $W\cdot T$.
This completes our discussion of the building blocks that we needed
to implement our construction.

Acknowledgements. This work is supported in part by the Austrian Research Funds
FWF (Schr"odinger fellowship 
J1530-TPH), NSF grant PHY-9511632, the Robert A. Welch Foundation and
by the Austrian National Bank under grant Nr. 6632.

\end{document}